\documentclass[12pt,preprint]{aastex}
\usepackage{rotating}
\usepackage{epsfig}
\usepackage{ifpdf}
\begin{document}

\title{The Proper Motions of the Double Radio Source n in the Orion BN/KL Region}

\author{
Luis F. Rodr\'\i guez\altaffilmark{1},
Sergio A. Dzib\altaffilmark{2}, Laurent Loinard\altaffilmark{1,2}, Luis Zapata\altaffilmark{1}, Laura G\'omez\altaffilmark{3,4,5}, Karl M. Menten\altaffilmark{2}, Susana Lizano\altaffilmark{1}}

\altaffiltext{1}{Instituto de Radioastronom\'\i a y Astrof\'\i sica, 
UNAM, Apdo. Postal 3-72 (Xangari), 58089 Morelia, Michoac\'an, M\'exico}
\altaffiltext{2}{Max Planck Institut  f\"ur Radioastronomie, Auf
dem H\"ugel 69, 53121 Bonn, Germany}
\altaffiltext{3}{Joint ALMA Observatory, Alonso de C\'ordoba 3107, Vitacura,
Santiago, Chile}
\altaffiltext{4}{CSIRO Astronomy and Space Science, PO Box 76,
NSW 1710 Epping, Australia}
\altaffiltext{5}{Departamento de Astronom\'\i a, Universidad de Chile,
Camino El Observatorio 1515, Las Condes, Santiago, Chile}

% \altaffiltext{3}{Instituto de Astrof\'\i sica de Andaluc\'\i a, CSIC, Glorieta de la Astronom\'\i a, s/n, E-18008, Granada, Spain}
\email{
l.rodriguez@crya.unam.mx}
 
\begin{abstract}
We have extended the time baseline for observations of the proper motions of radio sources in the
Orion BN/KL region from 14.7 to 22.5 years. We present improved determinations for the sources BN and I. In addition,
we address the proper motions of the double radio source n, that have been questioned in the literature. We confirm that all three sources
are moving away at transverse velocities of tens of km s$^{-1}$ from a region in-between them, where they were located about 500 years ago. Source n exhibits a new component that
we interpret as due to a one-sided ejection of free-free emitting plasma that took place after 2006.36. We used the highly accurate relative proper motions
between sources BN and I to determine that their closest separation took place in the year 1475$\pm$6, when they were within $\sim$100 AU or less from each other in the plane of the sky.
%Remarkably, while sources BN and I show
%clear ballistic (unaccelerated) motions, source n seems to be experiencing an acceleration whose cause...

\end{abstract}  

\keywords{Astrometry -- stars: formation --
stars: individual (BN, I, n) --
stars: radio continuum
}

\section{Introduction}

The Orion BN/KL region has long been known to exhibit the evidence of a violent explosive phenomenon that took place some 500 years ago (e.g. Bally \& Zinnecker 2005; G\'omez et al. 2008).
Filaments of molecular gas (the so-called Orion H$_2$ fingers; e.g. Bally et al. 2015; Youngblood et al. 2016) are moving away from a common origin at velocities in the range of 
tens to hundreds of km s$^{-1}$. The inner parts of this outflowing molecular gas have been imaged in CO by Zapata et al. (2009) and exhibit the same behavior.
Three compact radio sources, believed to trace stellar objects, are also known to recede from the same location at transverse
velocities in the range of 15 to 26 km s$^{-1}$ (G\'omez et al. 2008). The proper motions of these compact objects, known as sources BN, I and n, have
been studied also by Goddi et al. (2011), who confirmed the proper motions for sources BN and I previously determined by G\'omez et al. (2008) but 
noted that their measurements were consistent with a null proper motion for source n. In this paper we reanalyze the proper motions of these
compact radio sources taking advantage of the existence of new observations of high quality. The data discussed here cover the period from 1991.67 to 2014.17,
about 50\% larger than that used by G\'omez et al. (2008). Since the accuracy of the proper motion determinations improves with time as t$^{3/2}$
(Dzib et al. 2016) our results are about a factor of 2 more accurate than those reported in G\'omez et al. (2008).

\section{Observations}

To improve the determination of the proper motions of sources BN, I and n, we analyzed Very Large Array (VLA) and Jansky VLA archive observations 
made at 7 epochs between 1991.67 and 2014.17. In the case of the bright radio source BN we were able to add a previous observation of lower sensitivity made in
1985.05. These epochs were selected on the basis of a combination of the
best angular angular resolution and sensitivity possibles. In particular, all are made in the highest angular resolution A configuration. On the other hand, in the case of source n we did not take into account the 
observations of 2011.56 when the emission was seriously affected by the free-free emission from
a cloud of ionized gas (see below for a detailed discussion). The data were analyzed in the standard manner using the AIPS (Astronomical Imaging Processing System) and  the CASA (Common Astronomy
Software Applications) packages of NRAO, the former one for the VLA observations and the latter one for the Jansky VLA observations.  In the calibration stage we used
the pipeline provided for Jansky VLA observations by NRAO. The final analysis and production of contour images was all made within the older but more mature AIPS software. The detailed analysis
of a subset of the data is presented in G\'omez et al. (2008).

%Dzib et al. (2016) have discussed the parameters of these observations.
In Table 1 we show the proper motions determined for sources BN, I and n. The proper motions were calculated adding
in quadrature systematic errors of order $\sim0\rlap.{''}01$ to the formal position uncertainties to obtain a reduced $\chi^2$ = 1. The proper motions for
source BN and I coincide within about $\pm1$--$\sigma$ to $\pm2$--$\sigma$ with those of G\'omez et al. (2008). This result supports a ballistic (i.e. unaccelerated)
motion for these two sources. The interpretation of the proper motions of
source n will be discussed below.

\section{Individual Proper Motions}

\subsection{Sources BN and I}

In the bottom panels of Figures 1 and 2 we show the contour images of these two sources for epochs 1991.67 and 2012.76, where the displacement is
evident.
In the top panels of these figures we also show the positions of sources BN and I as a function of time. Clearly, the motions follow a straight line within the noise.
%The proper motions determined are consistent with those determined over smaller time baselines by G\'omez et al. (2008) and
%Goddi et al. (2011).

\begin{figure}
\centering
\vspace{-1.8cm}
\includegraphics[angle=-90,scale=0.4]{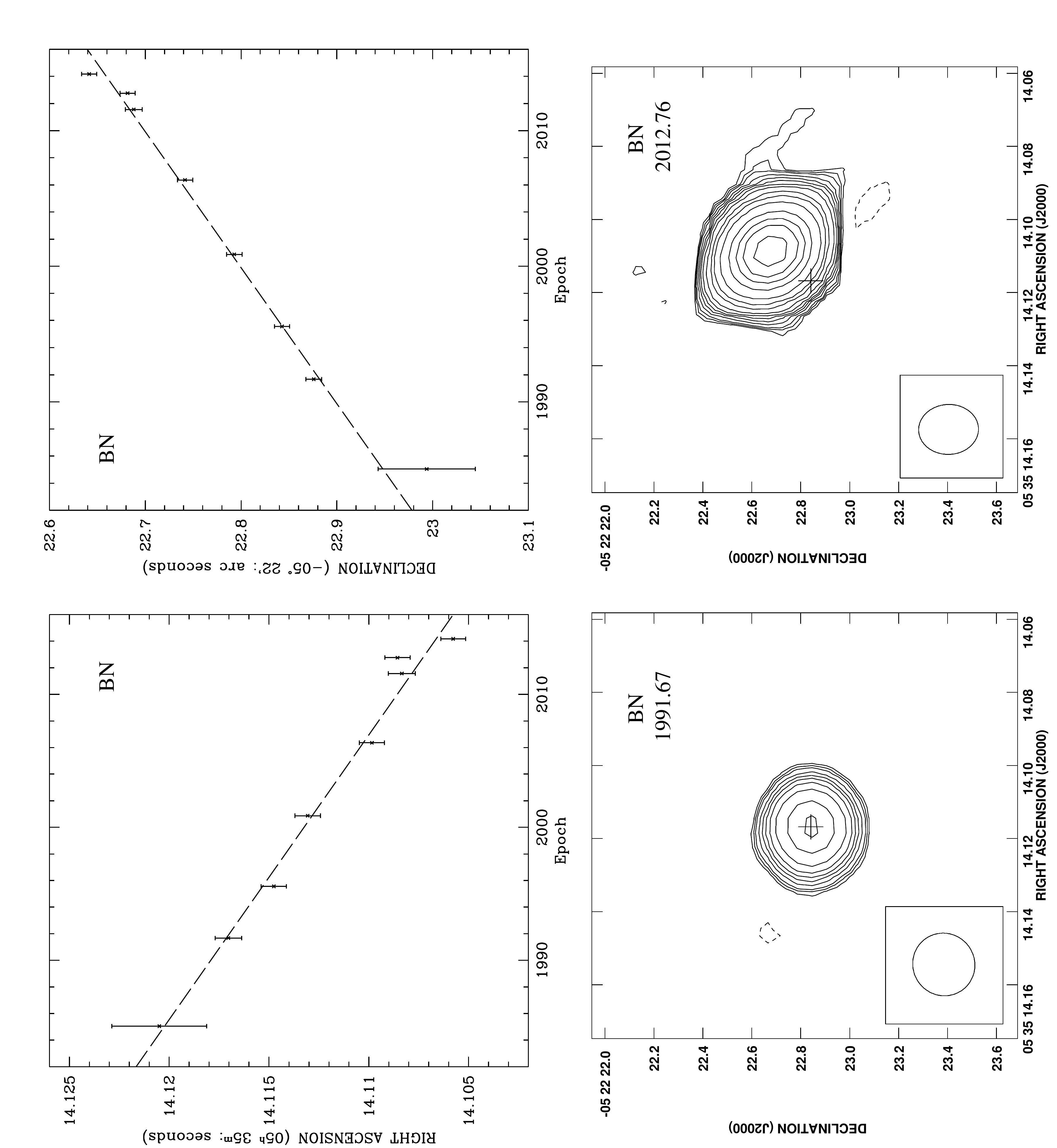}
\vskip -0.0cm
\caption{\small Top left: Proper motions in right ascension for source BN. Top right: Proper motions in declination for source BN.
The dotted line is a least-squares linear fit to the data. The resulting proper motions are given in Table 1.
Bottom left: VLA 8.4 GHz  continuum contour image of the source BN for epoch 1991.67.
The contours are -4, 4, 5, 6, 8, 10, 12, 15, 20, 30 and 40 times 88
$\mu$Jy beam$^{-1}$, the rms noise of the
image. 
The half-power contour of the synthesized beam  is shown in the bottom left corner
($0\rlap.{''}26 \times 0\rlap.{''}25;  PA = -55^\circ$). The cross marks the peak position of the emission for
epoch 1991.67. Bottom right: JVLA 7.3 GHz  continuum contour image of the source BN for epoch 2012.76
The contours are -4, 4, 5, 6, 8, 10, 12, 15, 20, 30, 40, 60, 100, 200, and 250 times 10
$\mu$Jy beam$^{-1}$, the rms noise of the
image. 
The half-power contour of the synthesized beam  is shown in the bottom left corner
($0\rlap.{''}24 \times 0\rlap.{''}20;  PA = 2^\circ$). The cross marks the peak position of the emission for
epoch 1991.67. The parameters of the fits to the proper motions are given in Table 1.
}
\label{fig1}
\end{figure}

\subsection{The Double Source n and the Associated Ejecta}

We now discuss the proper motions of the double source n. Source n appeared as double in the north-south direction for the three observations made in the period from 1991.67 to 2000.87 (see Figure 8 of G\'omez et al. 2008). We will refer to these components as
nN and nS. However, by 2006.36 the double source appeared unresolved, as a single elongated object (see Figure 8 of G\'omez et al. 2008).
In Figure 3 we show the contour images for epochs 1991.67 and 2012.76. The image for 1991.67 clearly shows the double morphology. The image for 2012.67 is complex and can be
described as constituted of two components, but much more separated and of different morphology than those
components seen in 1991.67. We interpret this image as follows. The northern component is taken to be the double source that remains spatially unresolved, as first observed in
2006.36. The southern component is interpreted as a cloud of ionized plasma that was ejected from one of the stars in the double system sometime after 2006.36.

The proper motions reported in Table 1 and shown in Figure 3 are obtained from fitting the double stellar component, with the position being the average of the double source, or the centroid of a
single elongated source when the double source was not resolved (which is the case for all images taken after 2006.36). This set of data contains two
new observations with respect to the data used by G\'omez et al. (2008). New observations over the following years are needed to test if the proper 
motions reported here predict the future position of the centroid of the
double stellar source.

If we assume that the ejection took place after 2006.36, we can estimate lower limits to the proper motion of the ejecta from the position shown in Figure 3. If we assume that it was ejected from the northern stellar component, we obtain a lower limit for the proper
motion of $\geq$110 mas yr$^{-1}$, while if it was ejected from the southern stellar component we obtain a lower limit for the proper
motion of $\geq$60 mas yr$^{-1}$. At a distance of 414 pc (Menten et al. 2007, see however a new estimate of 388 pc by Kounkel et al. 2016), these proper motions are equivalent to velocities in the plane of the sky of 220 km s$^{-1}$ and 120 km s$^{-1}$, 
respectively. These velocities are comparable to those observed in the proper motions of ejecta from young, intermediate-mass stars (e.g. Rodr{\'{\i}}guez-Kamenetzky et al. 2016).

The ejecta has to be dominantly one-sided (approximately to the south), because we do not see a major distortion in the northern component.
One-sided ejecta (sometimes called monopolar or asymmetric jets) 
from young stars is not an uncommon phenomenon and has been observed in sources such as OMC 1 South (Zapata et al. 2006), DG Tau (Rodr\'\i guez et al. 2012a), DG TauB (Rodr\'\i guez et al. 2012b),
HH~111 (G\'omez et al. 2013), NGC 1333-IRAS2A (Codella et al. 2014) and Serpens SMM1 (Hull et al. 2016).

\subsection{Testing the Plasma Ejection Scenario}

There are two ways to test this scenario. First, if the emission observed recently to the south of the double source is produced by ejected plasma, we expect the ejecta to become fainter on a recombination timescale.
From the 2012.76 observations (see Figure 3), we measure a flux density for the southern condensation of $S_{7.3~GHz}= 0.45 \, mJy$, 
and an angular size $ \theta_s^2 < (0.2 \, \arcsec)^2$. Assuming optically thin free-free emission, an electron temperature $T_e = 10^4$ K, the emission measure is
$ EM = n_e^2 l \gtrsim 5 \times 10^6 {cm^{-6} pc} \left[{S_\nu \over 0.45~ mJy}\right] \left[{\theta_s \over 0.2 \arcsec }\right]^{-2}.$
Assuming a characteristic scale $l/AU = (\theta_s/ \arcsec)( d / pc) \sim  83 \, AU $, where $d=414 $ pc is the distance
to Orion, the electron density of the plasma is 
$ n_e \gtrsim 10^5 \, cm^{-3} \left[{S_\nu \over 0.45 ~mJy}\right]^{1/2}  \left({ \theta_s \over 0.2 \arcsec }\right)^{-3/2}. $
The  ejected plasma will recombine in a timescale
$ t_{rec} = \left({n_e \alpha_2}\right)^{-1}  \lesssim 1 \, yr, $
where the recombination coefficient is $\alpha_2 = 2.6 \times 10^{-13}~ cm^3 ~s^{-1}$. 
Assuming a proper motion of the ejecta $\mu \sim$ 60 mas yr$^{-1}$, that corresponds in Orion to 120 km s$^{-1}$, the crossing time is
$ t_{cross} \sim {l / v}  \sim \, 3 \,  yr.  $ Because $t_{cross} >  t_{rec}$,
one expects the plasma to recombine as it moves away from the source\footnote{Nevertheless, if the 
ionized plasma is produced in a shock moving along the flow axis  (e.g., faster
wind reaching a slower upstream wind) the plasma would not recombine because it would be ionized in situ.}.

Furthermore, if the ionized plasma is expanding, the density will drop. If the expansion is isotropic, the plasma volume is
$V \propto l^3$ and the electron density will decrease like $n_e \propto l^{-3}$.  
For a jet condensation expanding with a fixed solid angle $\Omega$, at a distance $r$ from
the source, the volume is $V \propto \Omega r^2 \Delta r$. If the size $\Delta r \sim $constant, the volume will increase as $V \sim l^2$
and the electron density will decrease like $n_e \propto l^{-2}$. The
optically thin radio flux density is $S_\nu \propto  n_e^2 V$, and it will decrease like
$S_\nu \propto l^{-3}$ in the isotropic case, and like $S_\nu \propto l^{-2}$ in the case of the jet condensation. 
Both types of flux density decrease were observed in the thermal radio jet HH 80-81 by Mart\'\i ~ et al. (1998) who studied the evolution of
two jet condensations. In any case, one expects that the observed radio emission of the ejection would 
decrease very fast both due to recombination and expansion of the ionized gas.

The assumption that we are dealing with optically thin free-free emission in the ejecta is supported by an analysis of the spectral index of source n in epoch 2012.76.
Using the data of Forbrich et al. (2016) we obtain a total flux density of 1.45$\pm$0.03 mJy at 7.3 GHz and of 1.68$\pm$0.05 mJy at 4.7 GHz, This gives a spectral index
of -0.3$\pm$0.1, in agreement with Forbrich et al. (2016). For the southern component (the ejecta) we obtain flux densities of 0.45$\pm$0.02 and 0.49$\pm$0.03
at 7.3 and 4.7 GHz, respectively. This results in a spectral index of -0.2$\pm$0.2, consistent with optically thin free-free-emission. Finally, for the northern
component (proposed here to be the unresolved double stellar source) we obtain flux densities of 1.00$\pm$0.02 and 1.19$\pm$0.04
at 7.3 and 4.7 GHz, respectively. This gives a spectral index of -0.4$\pm$0.1, suggesting a possible non-thermal component.

The total flux density of source n as a function of time (Figure 4) shows an important increment in epoch 2011.56. We believe that this increase (that has
practically disappeared in more recent epochs) was the result of the ejection of the plasma cloud.

One can also estimate the mass-loss rate of an ionized jet,
$ \dot M_i = \Omega r^2 \mu m_H  n_e v \sim$
$ 3 \times 10^{-8} \left[{S_\nu \over 0.45 ~mJy}\right]^{1/2}
 \left[{\theta_s \over 0.2 \arcsec}\right]^{1/2} M_\odot ~yr^{-1}. $
Assuming a jet ionization fraction $x_e \sim 0.1$, one obtains a total mass-loss rate $\dot M \sim 10^{-7}~ M_\odot ~yr^{-1}$. This
implies that source n must have an accretion disk with a healthy mass accretion rate, one order of magnitude larger, 
$\dot M_{acc} \sim 10^{-6} M_\odot ~yr^{-1} $ (e.g., Shu et al. 1993).

Finally, Anglada et al. (2015) proposed a correlation between the radio luminosity of the jet, $S_{3.6cm} d^2$,  and the bolometric
luminosity of the source, $L_{bol}$  (their eq. [3]). Assuming a flat spectrum, this relation gives a  luminosity
$ L \sim 44 ~L_\odot $. From the observed IR fluxes from 5 - 10 $\mu m$  (Gezari et al. 1998), one obtains a lower limit 
to the luminosity of source n, $ L > 4 ~L_\odot$. %{\bf SL to get accretion luminosity need mass estimate...}

The second way of testing the scenario is to verify
that the proper motions derived for source n excluding the southern ejecta are consistent with the epoch of cluster disintegration, as determined from the well-behaved proper motions of
sources BN and I. To pursue this test, in Table 2 we give 
the proper motions corrected for the mean proper motions of the Trapezium-BN/KL region ($\mu_{\alpha} cos \delta$ = 1.07$\pm$0.09 mas/yr; $\mu_{\delta}$ = -0.84$\pm$0.16 mas/yr), as determined by Dzib et al. (2016). 
This relatively small correction is required to present the proper motions in the rest frame of Orion.
In Figure 5 we present the position of the sources BN, I and n as a function of time, as extrapolated from the proper motions in Table 2. The angle associated with each source represents the 2-$\sigma$ error in the proper motions.
We can see that the past positions of BN, I and n overlap in the past both in right ascension as in declination.
The data are consistent with a disintegration epoch during the XVth century with an initial position of $\alpha(J2000) = 05^h~35^m~14\rlap.^s41$;
$\delta(J2000) = -05^\circ~22'~28\rlap.{''}2$. The more accurate determination using only the relative positions of BN and I gives
a disintegration epoch of 1490$\pm$11 (G\'omez et al. 2008) and, with additional data points, of 1453$\pm$9 (Goddi et al. 2011). With more data available, we will repeat below this determination.

In Figure 6 we show the location of the
three sources and, with arrows, the proper motions in the frame of Orion for a period of 200 years.

\section{The Minimum Separation of BN and source I in the past}

As discussed by G\'omez et al. (2008), 
the minimum separation between BN and source I and when it took place in the past can be estimated accurately
using relative astrometry between these two sources. This determination is very accurate because the relative astrometry is not
affected by most of the error sources present in the absolute astrometry.

Following G\'omez et al. (2008) we define as $x(t)$ and $y(t)$ the separations in 
right ascension and declination of BN with respect to source I 
as a function of epoch $t$ and as $\mu_x$ and $\mu_y$  the proper motions in 
right ascension and declination. 

%. Assuming that their proper motions are
%!TEX encoding = UTF-8 Unicodelinear, the displacements will be given by

%$$x(t) = x(2000.0) + \mu_x~ (t-2000.0),$$

%\noindent and

%$$y(t) = y(2000.0) + \mu_y~ (t-2000.0),$$ 

%\noindent where $x(2000.0)$ and $y(2000.0)$ are the displacements
%in right ascension and declination for epoch 2000.0,
%and $\mu_x$ and $\mu_y$ are the proper motions in 
%right ascension and declination. 

%The separation of the sources as a function of time, $s(t)$, will then be
%given by:

%$$s(t) = (x^2(t) + y^2(t))^{1/2}.$$

%Differentiating and equating to 0, we find that
The minimum separation takes place at an epoch $t_{min}$ given by

$$t_{min} = -{{x(2000.0) \mu_x + y(2000.0) \mu_y} \over { \mu_x^2 + \mu_y^2}},$$ 

\noindent and that this minimum
separation is 

$$s_{min} = {{|x(2000.0) \mu_y - y(2000.0) \mu_x|} \over {(\mu_x^2 + \mu_y^2)^{1/2}}}.$$

The least squares fit to 23 data points (14 from G\'omez et al. 2008 and 9 from this paper) from the VLA, Jansky VLA and ALMA archives (shown in
Table 3 and Figure 7) gives
%$x(2000.0) = -5.9385 \pm 0.0025$ arcsec, $\mu_x = -0.01136\pm 0.00023$ arcsec yr$^{-1}$,
%$y(2000.0) = 7.7292 \pm 0.0025$ arcsec, and $\mu_y = 0.01483 \pm 0.00024$ arcsec yr$^{-1}$.

$x(2000.0) = -5.9385 \pm 0.0024$ arcsec, $\mu_x = -0.01133\pm 0.00022$ arcsec yr$^{-1}$,

$y(2000.0) = 7.7282 \pm 0.0024$ arcsec, $\mu_y = 0.01472 \pm 0.00022$ arcsec yr$^{-1}$.

With these values, we obtain:

$$t_{min} = 1475 \pm 6,$$

$$s_{min} = 0\rlap.{''}007 \pm 0\rlap.{''}093,$$

\noindent where the errors were calculated 
using standard propagation error theory (Wall \& Jenkins 2003).
The inclusion in this analysis of observations of different configurations and frequencies (and even different radio telescopes, one data point
comes from ALMA) can affect the results. We argue, however, that both BN and source I are quite compact and that the determination of their centroids is not greatly affected
by angular resolution, except in the sense that the precision of the astrometrical positions depends linearly with angular resolution but that this effect should
not produce systematic shifts. As far as it is known, the soures do not present optical depth effects that could produce frequency-dependent displacements in the positions. Finally,
as already noted above, the astrometry of this section is relative and this minimizes position shifts produced by the use of different complex gain calibrators.\rm 
We conclude that, about 540 years ago,
BN and source I were within less than $\sim0\rlap.{''}28$ (3-$\sigma$ upper limit) from each other
in the plane of the sky. At a distance of 414 pc this corresponds to a physical separation of $\sim$120 AU. We conclude that if by that epoch BN and/or source I
had protoplanetary disks of typical dimensions (i.e. 100 AU) the gas in these disks must have suffered a major disruption and could have originated the explosive molecular
outflow present in the region. This result, however, poses a difficulty for the cluster disruption scenario as recently
stressed by  Plambeck \& Wright (2016). These authors show, from ALMA observations, that source I has at present a 
disk with mass in the range of 0.02--0.2 $M_\odot$ and a diameter of $\sim$100 AU. They note that such a massive disk could not have formed in only 500 years
from  Bondi-Hoyle  accretion  (gravitational
focusing) as source I moves in the dense ($\sim$10$^{7}$ cm$^{-3}$) ambient medium of the BN/KL region. However, Goddi et al. (2011) have pointed out that
if source I existed as a softer binary before the close encounter
this could have enabled preservation of the original accretion disk, although truncated to its present radius of $\sim$50 AU. 
The numerical analysis of Moeckel \& Goddi (2012) confirms to first order the plausibility of the scattering scenario of Goddi et al. (2011)
and suggests that the original disk was largely preserved.
There is still much to understand of the explosive
phenomena present in BN/KL.

%A possible explanation is that the star is associated with component nN. while component nS is tracing ejecta from the star. A fresh ejection to
%the south would appear as an accelerating component. 

%If we consider acceleration, does ejection time of source n becomes consistent with those of BN and I.
\pagebreak

\begin{figure}
\centering
\vspace{-1.8cm}
\includegraphics[angle=-90,scale=0.4]{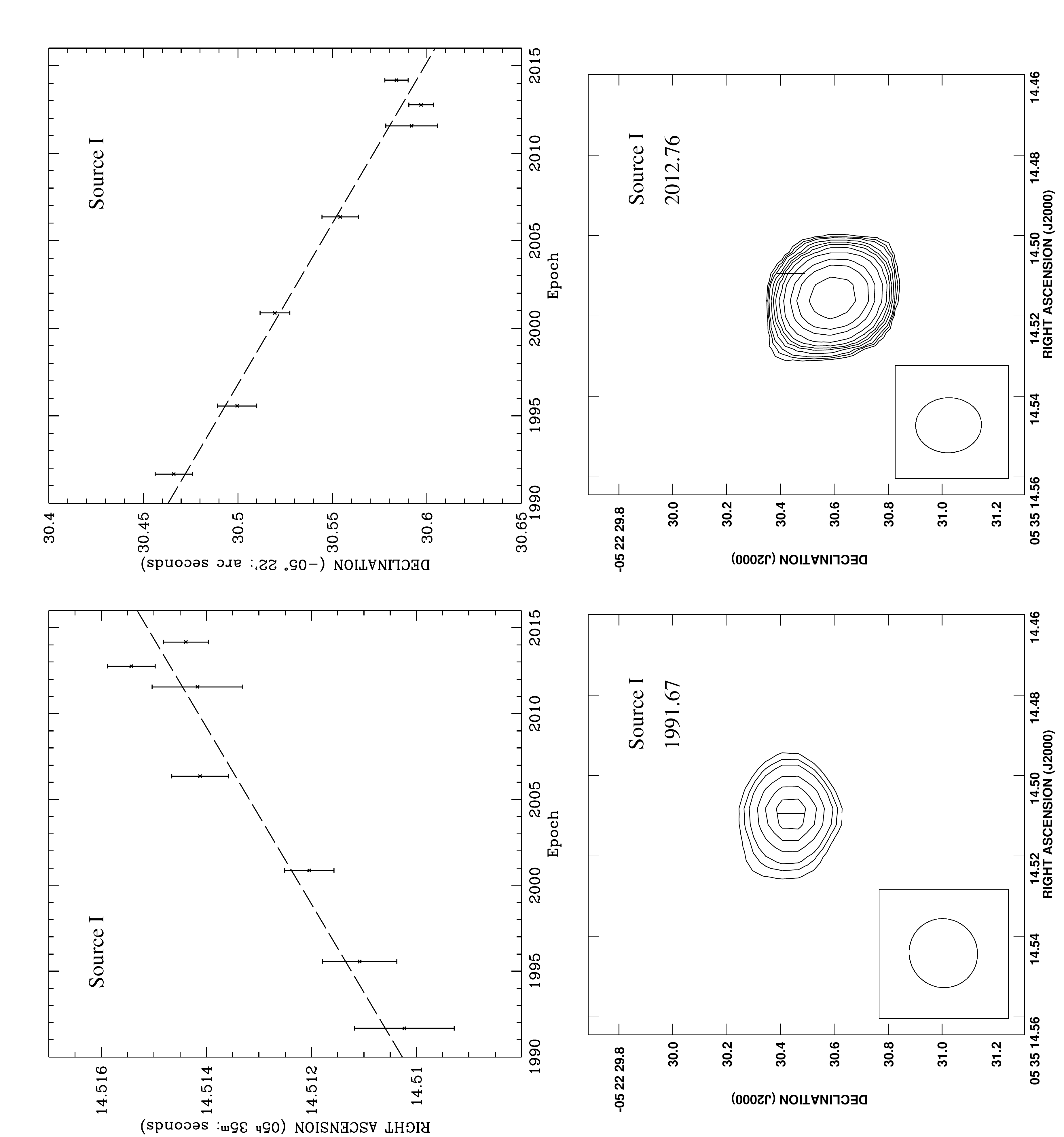}
\vskip-0.0cm
\caption{\small Same as in Figure 1 but for source I.
}
\label{fig2}
\end{figure}

\pagebreak

\begin{figure}
\centering
\vspace{-1.8cm}
\includegraphics[angle=-90,scale=0.4]{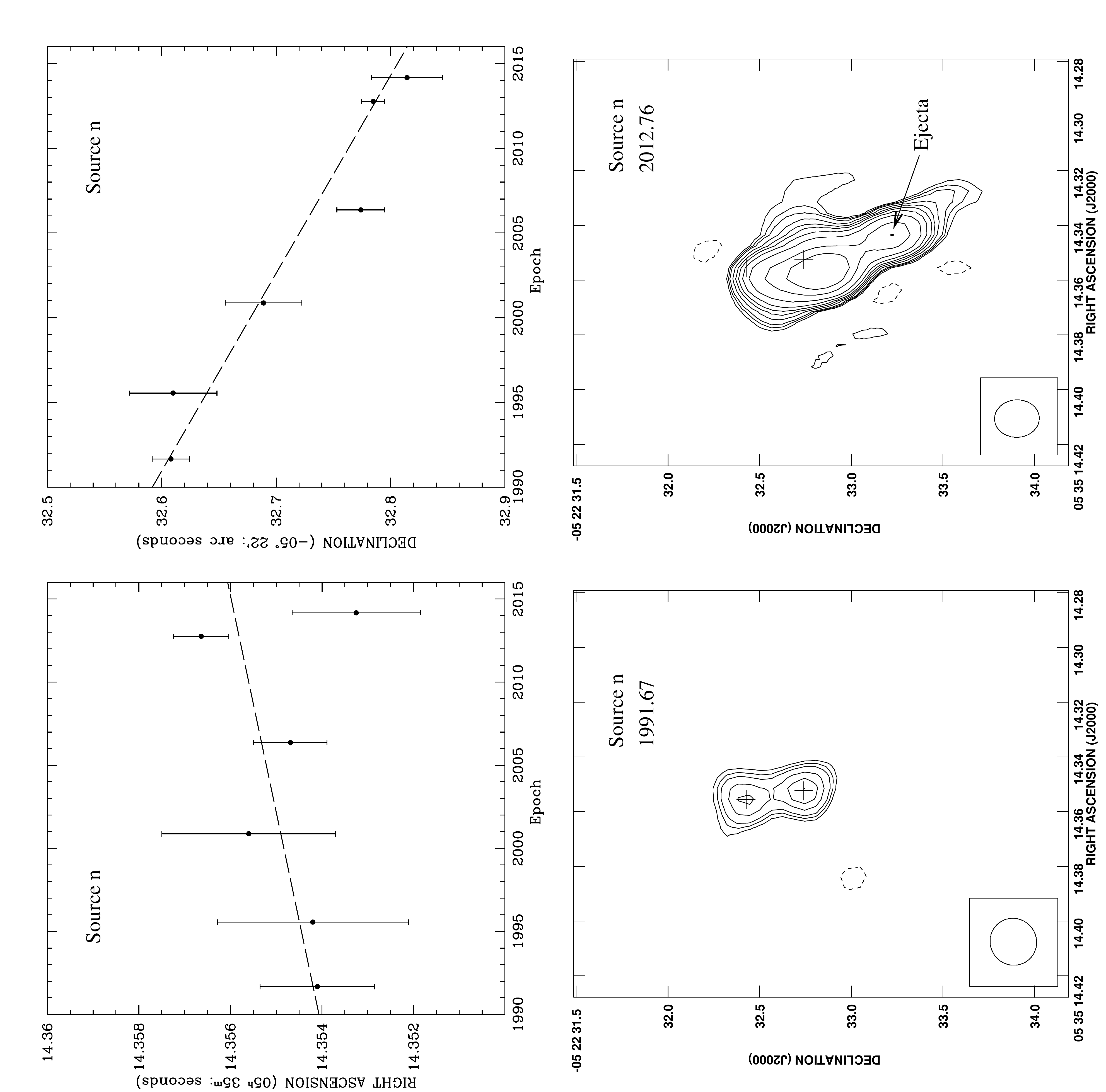}
\vskip-0.0cm
\caption{\small Same as in Figure 1 but for source n. In the image of 2012.76 the arrow indicates the component that
is interpreted as a recent ionized ejecta.
}
\label{fig3}
\end{figure}

\pagebreak

\begin{deluxetable}{c c c c c c c c }
\tabletypesize{\scriptsize}
\tablecaption{Parameters of Radio Sources in the Orion BN/KL Region}
\tablehead{                      
\colhead{                        }     &
\multicolumn{2}{c}{Position$^a$} &
\colhead{Flux Density$^b$}                     &   
\multicolumn{2}{c}{Proper Motions$^c$} &
\colhead{Total Proper Motion$^d$}                     \\
\colhead{Name}                              &
\colhead{$\alpha$(J2000)}    &
\colhead{$\delta$(J2000)}    &
\colhead{(mJy)}                     &
\colhead{$\mu_{\alpha} cos \delta$}                     &
\colhead{$\mu_{\delta}$}                     &
\colhead{Magnitude; PA}                     \
}
\startdata

BN & $05^h~35^m~14\rlap.^s110$ & $-05^\circ~22{'}~22\rlap.{''}67$ & 4.06$\pm$0.20  & --7.0$\pm$0.4 & +10.0$\pm$0.4 & 12.2$\pm$0.4; --35$\pm$2 \\
%nS & $05^h~35^m~14\rlap.^s343$ & $-05^\circ~22{'}~33\rlap.{''}23$ &   0.72$\pm$0.04 &  -9.1$\pm$1.7 &  -25.8$\pm$2.6 & 27.4$\pm$2.5; -161$\pm$5 \\
n & $05^h~35^m~14\rlap.^s357$ & $-05^\circ~22{'}~32\rlap.{''}78$ & 1.45$\pm$0.03  &  +1.1$\pm$0.9 & --8.6$\pm$0.6 & 8.7$\pm$0.6; +173$\pm$4 \\
I & $05^h~35^m~14\rlap.^s516$ & $-05^\circ~22{'}~30\rlap.{''}59$ & 0.98$\pm$0.05  &  +2.9$\pm$0.4 & --5.4$\pm$0.4 & 6.1$\pm$0.4; +152$\pm$4 \\

\enddata
\tablecomments{
(a): Equinox J2000.0 and epoch 2012.76.
(b): At the frequency of 7.3 GHz.
(c): In mas/yr.
(d): Magnitude in mas/yr, PA in degrees.}
\end{deluxetable}

\pagebreak

\pagebreak

\begin{deluxetable}{c c c c c}
\tabletypesize{\scriptsize}
\tablecaption{Proper Motions Corrected to the Orion Frame and Deconvolved Dimensions of the Radio Sources}
\tablehead{  
 \colhead{}    &         
\multicolumn{2}{c}{Proper Motions$^a$} &
\colhead{Total Proper Motion$^b$}   &
\colhead{Deconvolved Dimensions$^c$} \\         
\colhead{Name}                              &
\colhead{$\mu_{\alpha} cos \delta$}                     &
\colhead{$\mu_{\delta}$}                     &
\colhead{Magnitude; PA}            &
\colhead{($\theta_{maj}\times \theta_{min}; PA$)} \
}
\startdata

BN &  --8.1$\pm$0.4 & +10.8$\pm$0.4 & 13.5$\pm$0.4; --37$\pm$2  & $0\rlap.{''}19\pm0\rlap.{''}01 \times 0\rlap.{''}07 \pm 0\rlap.{''}01; +58^\circ \pm 1^\circ$ \\
%n(1991-2006) & +2.0$\pm$2.3 &  -13.6$\pm$3.1 & 13.7$\pm$3.1; +172$\pm$13 & +1.0$\pm$2.3 &  -12.8$\pm$3.1 & 12.8$\pm$3.1; +176$\pm$14 \\
%n(2011-2014) & -16.0$\pm$11.8 & -56.6$\pm$15.8 & 58.8$\pm$15.5; -164$\pm$15  & -17.0$\pm$11.8 & -55.8$\pm$15.8 & 58.3$\pm$15.5; -163$\pm$15 \\
n & +0.0$\pm$0.9 & --7.8$\pm$0.6 & 7.8$\pm$0.6; +180$\pm$4  & $0\rlap.{''}43\pm0\rlap.{''}01 \times 0\rlap.{''}17 \pm 0\rlap.{''}01; +16^\circ \pm 1^\circ$  \\ 
I & +1.8$\pm$0.4 & --4.6$\pm$0.4 & 4.9$\pm$0.4; +159$\pm$5  & $0\rlap.{''}13\pm0\rlap.{''}01 \times \leq0\rlap.{''}03 \pm 0\rlap.{''}01; +58^\circ \pm 3^\circ$  \\

\enddata
\tablecomments{
(a): In mas/yr.
(b): Magnitude in mas/yr, PA in degrees.
(c): From the data of the Jansky VLA project SD0630 observed in epoch 2012.76 at the frequency of 7.3 GHz.}
\end{deluxetable}

\clearpage

\begin{deluxetable}{lccccc}
\tabletypesize{\scriptsize}
% \tablewidth{18.5cm}
\tablecaption{VLA, Jansky VLA and ALMA Data Used for the Determination of the Relative Proper Motions
between BN and source I}
%\small
\tablehead{
\colhead{}  & \colhead{}  & \colhead{$\lambda$} 
& \colhead{Synthesized Beam} & $\Delta \alpha$\tablenotemark{c} & 
$\Delta \delta$\tablenotemark{c} \\  
\colhead{Epoch\tablenotemark{a}}  & \colhead{Project} & \colhead{(cm)}
& \colhead{($\theta_M \times \theta_m; PA$)\tablenotemark{b}} & (seconds) & (arcsecs) \\
}
\startdata
\bf 1985 Jan 19 (1985.05) \rm & AM143 & 1.3 & $0\rlap.{''}10\times0\rlap.{''}09;-7^\circ$ &
-0.38540$\pm$0.00079 & 7.5123$\pm$0.0128  \\
1985 Jan 19 (1985.05) & AM143 & 2.0 & $0\rlap.{''}16\times0\rlap.{''}13;+32^\circ$ & 
-0.38550$\pm$0.00074 & 7.5105$\pm$0.0110  \\
1986 Apr 28 (1986.32) & AC146 & 2.0 & $0\rlap.{''}15\times0\rlap.{''}14;+10^\circ$ &
-0.38678$\pm$0.00096 & 7.5202$\pm$0.0133  \\
\bf 1991 Sep 02 (1991.67) \rm & AM335 & 1.3 & $0\rlap.{''}10\times0\rlap.{''}10;-16^\circ$&
-0.39143$\pm$0.00074 & 7.5919$\pm$0.0110  \\
1994 Apr 29 (1994.33) & AM442 & 3.6 & $0\rlap.{''}23\times0\rlap.{''}20;+1^\circ$ &
-0.39408$\pm$0.00070 & 7.6425$\pm$0.0108  \\
\bf 1995 Jul 22 (1995.56) \rm & AM494 & 3.6 & $0\rlap.{''}26\times0\rlap.{''}32;+32^\circ$ &
-0.39422$\pm$0.00079 & 7.6768$\pm$0.0116  \\
1996 Nov 21 (1996.89) & AM543 & 3.6 & $0\rlap.{''}32\times0\rlap.{''}25;+0^\circ$ &
-0.39541$\pm$0.00073 & 7.6988$\pm$0.0117  \\
1997 Jan 11 (1997.03) & AM543 & 3.6 & $0\rlap.{''}33\times0\rlap.{''}25;-10^\circ$ &
-0.39471$\pm$0.00078 & 7.7066$\pm$0.0122  \\
2000 Nov 10 (2000.86) & AM668 & 0.7 & $0\rlap.{''}06\times0\rlap.{''}05;-25^\circ$ &
-0.39839$\pm$0.00068 & 7.7309$\pm$0.0102  \\
\bf 2000 Nov 13 (2000.87) \rm & AM668 & 3.6 & $0\rlap.{''}24\times0\rlap.{''}22;+3^\circ$ &
-0.39898$\pm$0.00078 & 7.7282$\pm$0.0121  \\
2002 Mar 31 (2002.25) & AG622 & 0.7 & $0\rlap.{''}05\times0\rlap.{''}03;+25^\circ$ &
-0.40002$\pm$0.00079 & 7.7541$\pm$0.0131  \\
2004 Nov 06 (2004.85) & AB1135 & 3.6 & $0\rlap.{''}23\times0\rlap.{''}20;+1^\circ$ &
-0.40249$\pm$0.00077 & 7.8079$\pm$0.0120  \\
\bf 2006 May 12 (2006.36) \rm & AR593 & 3.6 & $0\rlap.{''}26\times0\rlap.{''}22;-2^\circ$ &
-0.40426$\pm$0.00077 & 7.8142$\pm$0.0119  \\
2007 Dec 14 (2007.95) & AR635 & 0.7 & $0\rlap.{''}21\times0\rlap.{''}18;+20^\circ$ &
-0.40410$\pm$0.00069 & 7.8365$\pm$0.0105  \\
2009 Jan 12 (2009.03) & AC952 & 0.7 & $0\rlap.{''}08\times0\rlap.{''}05; +7^\circ$ &
-0.40511$\pm$0.00063 & 7.8500$\pm$0.0127 \\
2011 Jun 04 (2011.42) & 10B-175 & 0.9 & $0\rlap.{''}14\times0\rlap.{''}09; -71^\circ$ &
-0.40618$\pm$0.00031 & 7.8948$\pm$0.0036 \\
2011 Jul 02 (2011.50) & BL175 & 6.0 & $0\rlap.{''}30\times0\rlap.{''}27;+46^\circ$ &
-0.40880$\pm$0.00297 & 7.8981$\pm$0.0243 \\
\bf  2011 Jul 24 (2011.56)  \rm & BL175 & 6.0 & $0\rlap.{''}47\times0\rlap.{''}24; -47^\circ$ &
 -0.40779$\pm$0.00250  &  7.9007$\pm$0.0305 \\
 2011 Aug 29 (2011.66) & BL175 & 6.0 & $0\rlap.{''}33\times0\rlap.{''}25; -30^\circ$ &
 -0.40583$\pm$0.00086  &  7.9039$\pm$0.0139 \\
\bf  2012 Oct 03 (2012.76) \rm & SD0630 & 4.0 & $0\rlap.{''}22\times0\rlap.{''}20; -7^\circ$ &
 -0.40687$\pm$0.00038   & 7.9154$\pm$0.0058 \\
2014 Mar 03 (2014.17) & 13B-085 & 3.3 & $0\rlap.{''}25\times0\rlap.{''}18; -38^\circ$ &
 -0.40658$\pm$0.00066  &  7.9521$\pm$0.0135  \\
 \bf 2014 Mar 03 (2014.17)  \rm & 13B-085 & 5.5 & $0\rlap.{''}44\times0\rlap.{''}29; -38^\circ$ &
 -0.40862$\pm$0.00033 & 7.9425$\pm$0.0053 \\
  2014 Jul 14 (2014.57)\tablenotemark{d}  & 2012.1.00123.S & 0.09 & $0\rlap.{''}25\times0\rlap.{''}18; +66^\circ$ &
 -0.40763$\pm$0.00111 & 7.93152$\pm$0.0136 \\
\enddata
\tablenotetext{a}{The epochs used for the determination of the proper motions shown in Figures 1, 2 and 3 are indicated in boldface.}
\tablenotetext{b}{Major axis$\times$minor axis in arcsec; PA in degrees.}
\tablenotetext{c}{Positional offsets of BN with respect to source I in
right ascension and declination.}
\tablenotetext{d}{Data from the Atacama Large Millimeter Array (ALMA).}
%% You can append references to a table using the \tablerefs command.
\end{deluxetable}

\clearpage

\begin{figure}
\centering
\vspace{-1.8cm}
\includegraphics[angle=0,scale=0.6]{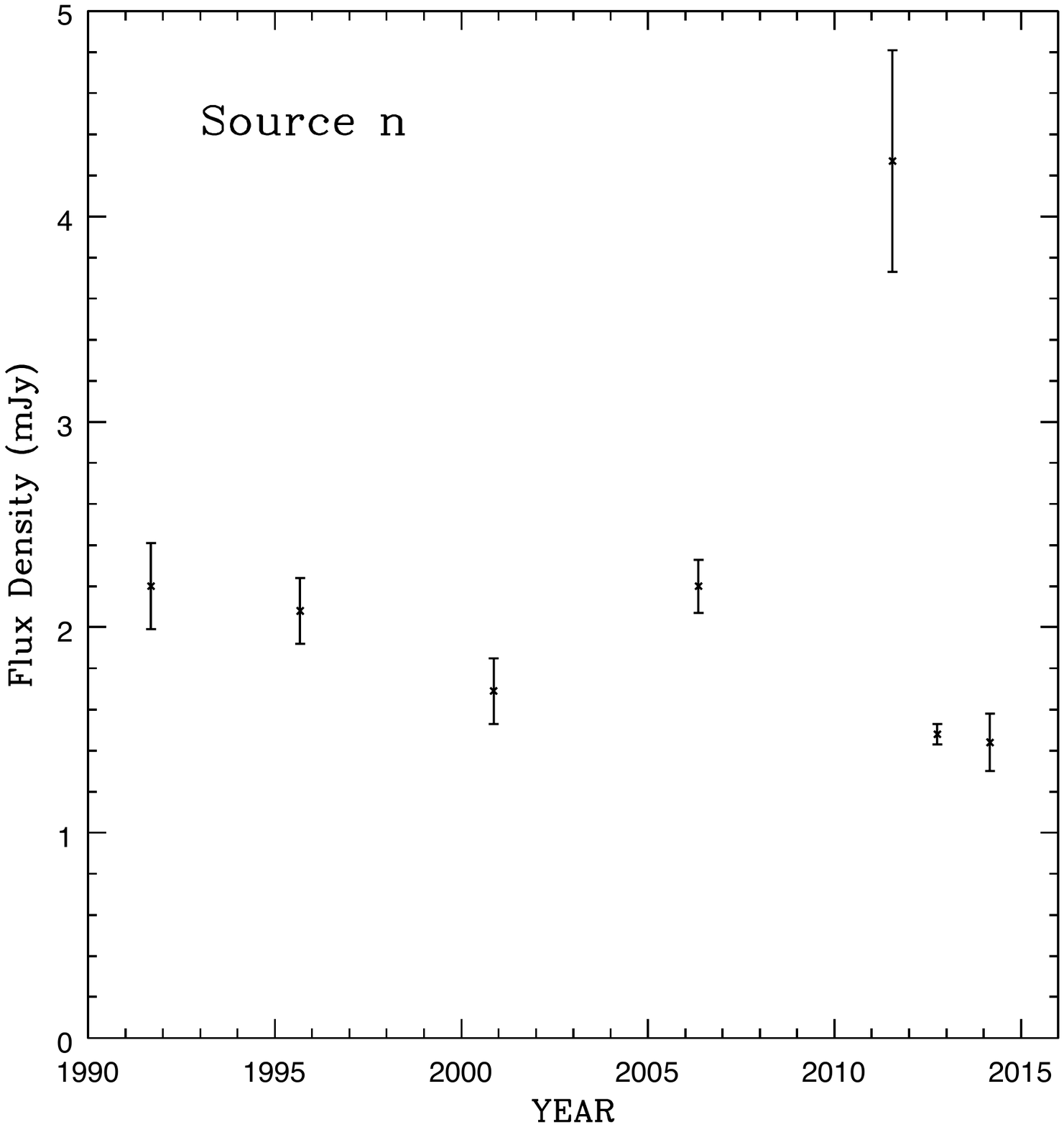}
\vskip-0.0cm
\caption{\small 8.4 GHz flux density for source n as a function of time. The increase for epoch 2011.56 is proposed to be associated with the
ejection. These flux densities have been obtained at somewhat different frequencies and have all been normalized to a frequency of 8.4 GHz
following the results of Forbrich et al. (2016) that imply that the frequency dependence of the centimeter flux density of source n is given by $S_\nu \propto \nu^{-0.3}$.
}
\label{fig4}
\end{figure}

\pagebreak

\begin{figure}
\centering
\vspace{-2.0cm}
\includegraphics[angle=-90,scale=0.50]{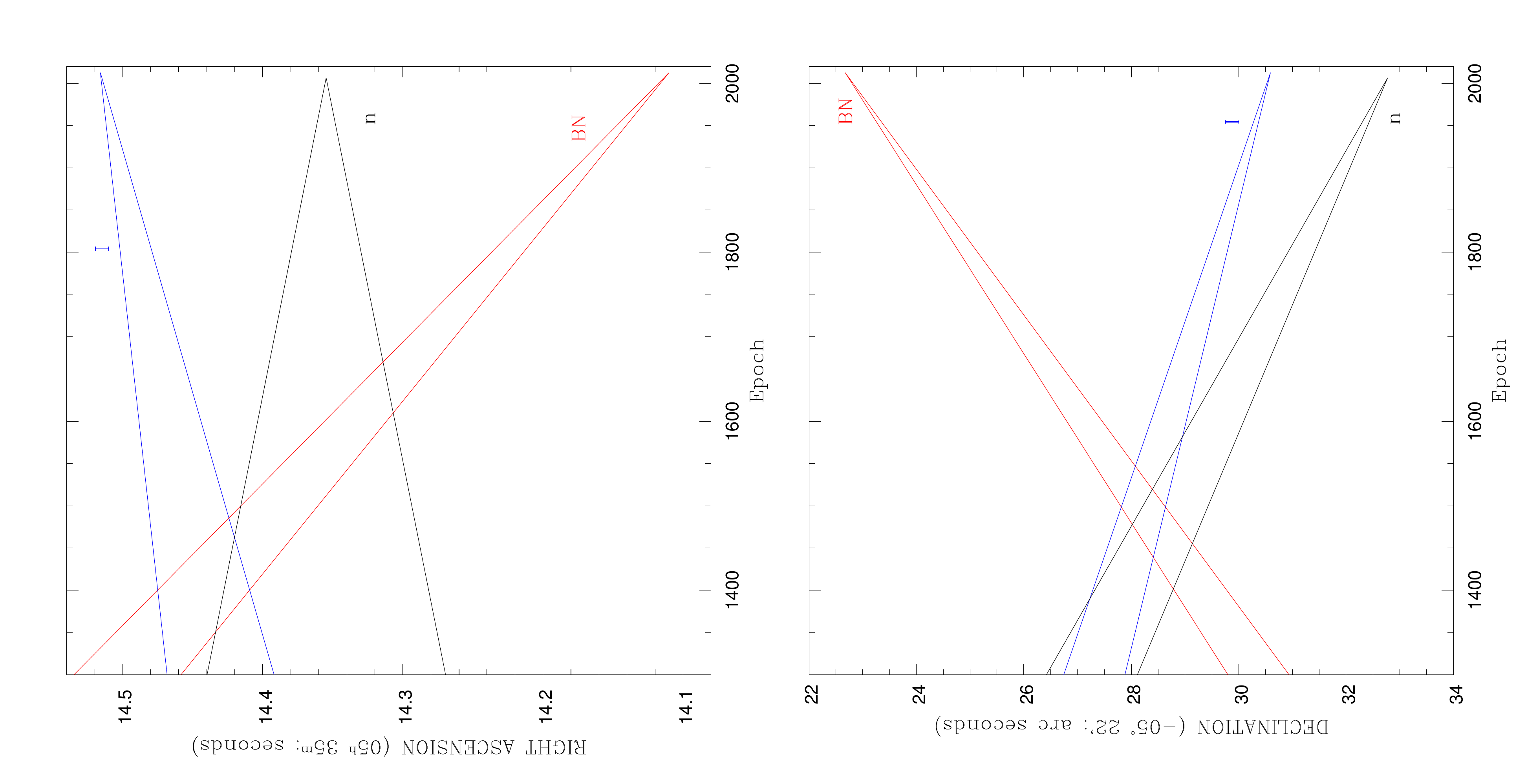}
\vskip-0.0cm
\caption{\small Proper motions of the sources BN (red), I (blue) and n (black) in the rest frame of Orion. The angles represent the $\pm$2-$\sigma$ error range in proper motions.
The data are consistent with a disintegration epoch in the 1400-1500 period with an initial position of $\alpha(J2000) = 05^h~35^m~14\rlap.^s41$;
$\delta(J2000) = -05^\circ~22'~28\rlap.{''}2$.
}
\label{fig5}
\end{figure}

\begin{figure}
\centering
\vspace{-1.5cm}
\includegraphics[angle=0,scale=0.35]{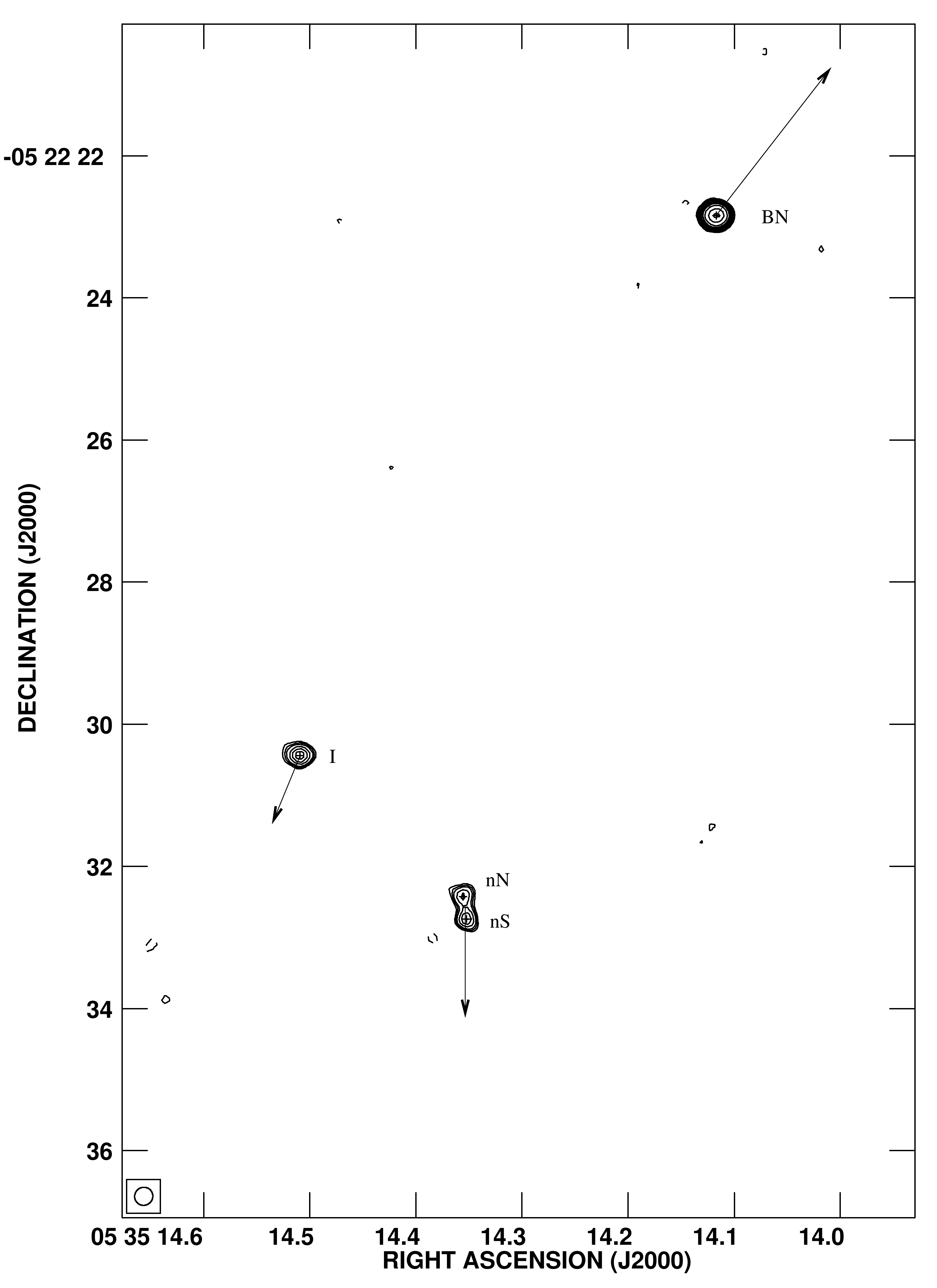}
\vskip-0.0cm
\caption{\small VLA 8.4 GHz  continuum contour image of the BN/KL region for epoch 1991.67.
The contours are -4, 4, 5, 6, 8, 10, 12, 15, 20, 30 and 40 times 88
$\mu$Jy beam$^{-1}$, the rms noise of the
image. The radio sources in the region are identified and the arrows indicate the proper motions in the frame of Orion for
a period of 200 years. 
The half-power contour of the synthesized beam  is shown in the bottom left corner
($0\rlap.{''}26 \times 0\rlap.{''}25;  PA = -55^\circ$).
}
\label{fig6}
\end{figure}

\begin{figure}
\centering
\vspace{-1.5cm}
\includegraphics[angle=-90,scale=0.55]{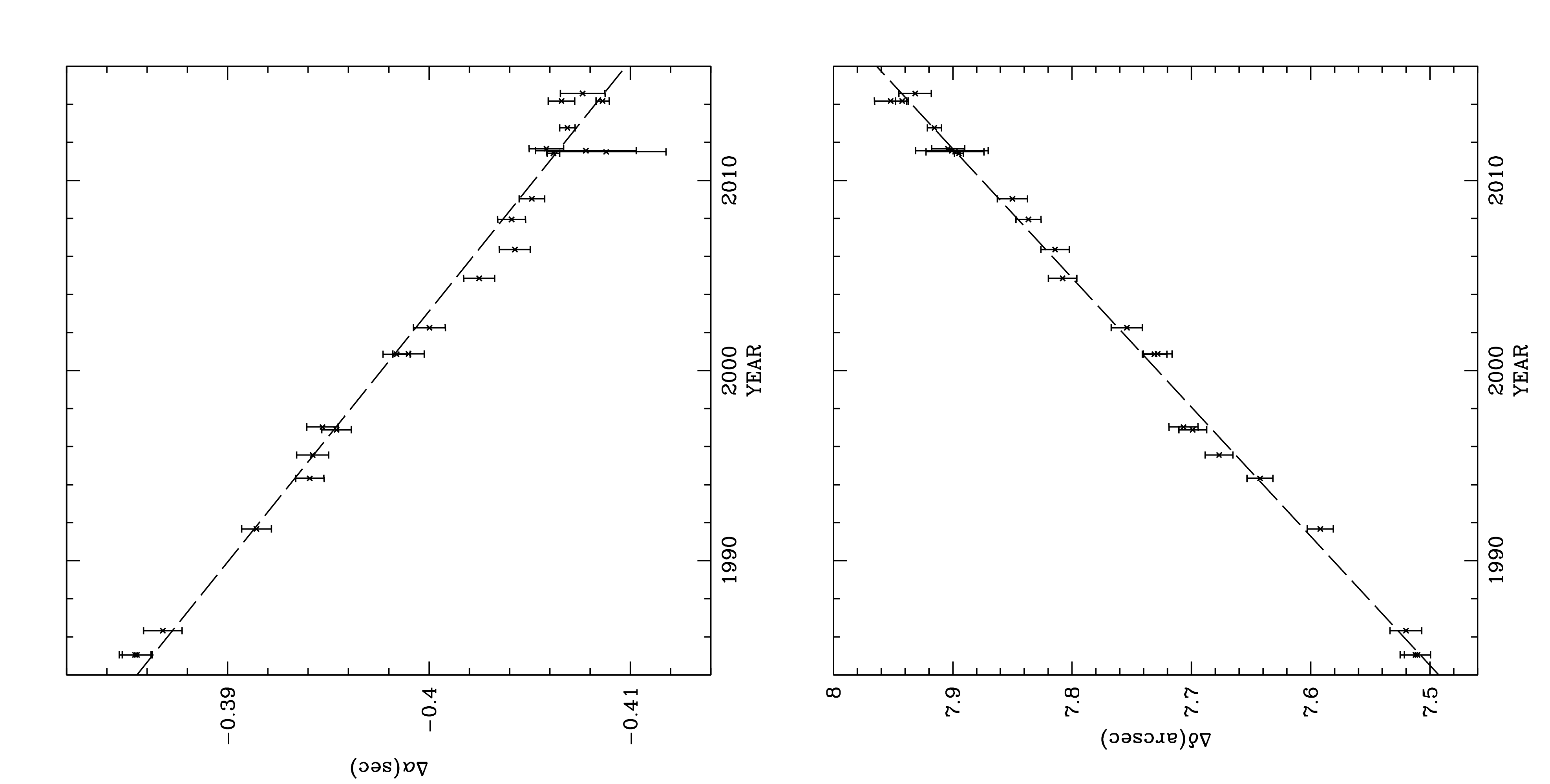}
\vskip-0.5cm
\caption{\small Proper motions of BN with respect to source I for right ascension (top) and declination (bottom).
The dashed lines represent the least-squares fits to the data.
}
\label{fig7}
\end{figure}

\pagebreak

\section{Conclusions}

The analysis of archive VLA data as well as new high sensitivity Jansky VLA observations allowed a new improved determination of the proper
motions of the stellar sources BN, I and n in the Orion BN/KL region. 
The proper motions of sources BN and I are consistent with previous measurements in the literature and a ballistic nature.
The proper motions of the double source n are complicated by the appearance of a new component, most likely a cloud of ionized gas that was ejected
from one of the stars in source n after 2006.36. We estimate that this cloud is moving at velocities of more than 120 km s$^{-1}$ in the plane of the sky.
The proper motions of BN, I and n are consistent with a disintegration epoch in the XV century.
Finally, we used the relative proper motions
between sources BN and I to accurately determine that the closest separation took place in the year 1475$\pm$6, when they were within 110 AU or less from each other in the plane of the sky.

\acknowledgments

This research has made use of the SIMBAD database,
operated at CDS, Strasbourg, France.
LFR, LZ and LL are grateful to CONACyT, Mexico and DGAPA, UNAM for their financial
support. SL acknowledges support  from project DGAPA IN105815 and CONACytT 238631.
This paper makes use of the following ALMA data: ADS/JAO.ALMA\#2012.1.00123.S. ALMA is a partnership of ESO (representing its 
member states), NSF (USA) and NINS (Japan), together with NRC (Canada) and NSC and ASIAA (Taiwan), in cooperation with the Republic of Chile. The Joint ALMA Observatory is operated by ESO, AUI/NRAO and NAOJ.

\clearpage

\end{document}